\documentclass[a4paper,fleqn,usenatbib]{mnras}

\usepackage[T1]{fontenc}
\usepackage{ae,aecompl}
\usepackage[symbol]{footmisc}

\usepackage{graphicx}	
\usepackage{amsmath}	
\usepackage{amssymb}	
\usepackage{hyperref}

\newcommand{\pcm}{\ensuremath{\mathrm{\, cm^{-2}}}}	
\newcommand{\msun}{\ensuremath{\mathrm{M_{\odot}}}}
\newcommand{\ten}[1]{\times 10^{#1}}
\newcommand{\sub}[2]{\ensuremath{#1_{\mathrm{#2}}}}

\newcommand{\saxj}{SAX J1808.4-3658}
\newcommand{\gs}{GS 1826-24}

\newcommand{\mdot}{\ensuremath{\dot{M}}}
\newcommand{\mdotavg}{\ensuremath{\langle \mdot \rangle}}
\newcommand{\dtavg}{\ensuremath{\langle \Delta t \rangle}}
\newcommand{\kepler}{\textsc{kepler}}
\newcommand{\lunits}{\ensuremath{\mathrm{\, erg \, s^{-1}}}}

\newcommand{\flunits}{\ensuremath{\mathrm{\, erg \pcm}}}

\newcommand{\dt}[1]{\ensuremath{\Delta t_\mathrm{#1}}}
\newcommand{\Eb}{{E}_\mathrm{b}}

\newcommand{\cno}{\ensuremath{Z_\mathrm{CNO}}}
\newcommand{\hyd}{\ensuremath{X_0}}
\newcommand{\hel}{\ensuremath{Y_0}}

\defcitealias{galloway_helium-rich_2006}{G06}

\title[Simulating X-ray bursts during an accretion event]{Simulating X-ray bursts during a transient accretion event}

\author[Z. Johnston et al.]{
Zac Johnston,$^{1,2,3}$\thanks{E-mail: zac.johnston@monash.edu}
Alexander Heger,$^{1,2,3,4,5}$
Duncan K. Galloway$^{1,2,3}$
\\
$^{1}$School of Physics and Astronomy, Monash University, Victoria 3800, Australia\\
$^{2}$Monash Centre for Astrophysics (MoCA), Monash University, Victoria 3800, Australia\\
$^{3}$Joint Institute for Nuclear Astrophysics - Center for the Evolution of the Elements, East Lansing, Michigan 48824, USA\\
$^{4}$Department of Astronomy,
Shanghai Jiao Tong University, Shanghai 200240, China\\
$^{5}$School of Physics and Astronomy, University of Minnesota, Minneapolis, 
Minnesota 55455, USA\\
}

\date{Accepted XXX. Received YYY; in original form ZZZ}

\pubyear{2017}

\begin{document}
\label{firstpage}
\pagerange{\pageref{firstpage}--\pageref{lastpage}}
\maketitle

\begin{abstract}
 Modelling of thermonuclear X-ray bursts on accreting neutron stars has to date focused on stable accretion rates. However, bursts are also observed during episodes of transient accretion. During such events, the accretion rate can evolve significantly between bursts, and this regime provides a unique test for burst models. The accretion-powered millisecond pulsar \saxj{} exhibits accretion outbursts every $2-3$ years. During the well-sampled month-long outburst of 2002 October, four helium-rich X-ray bursts were observed. Using this event as a test case, we present the first multi-zone simulations of X-ray bursts under a time-dependent accretion rate. We investigate the effect of using a time-dependent accretion rate in comparison to constant, averaged rates. Initial results suggest that using a constant, average accretion rate between bursts may underestimate the recurrence time when the accretion rate is decreasing, and overestimate it when the accretion rate is increasing. Our model, with an accreted hydrogen fraction of $X=0.44$ and a CNO metallicity of $Z_\mathrm{CNO}=0.02$, reproduces the observed burst arrival times and fluences with root mean square (RMS) errors of $2.8\,\mathrm{h}$, and $0.11\ten{-6} \flunits$,  respectively. Our results support previous modelling that predicted two unobserved bursts, and indicate that additional bursts were also missed by observations.
\end{abstract}




\section{Introduction}
Type I X-ray bursts are thermonuclear flashes in the accreted envelopes of neutron stars \citep{belian_discovery_1976, grindlay_discovery_1976}. In low-mass X-ray binaries (LMXBs), a mix of hydrogen and helium is transferred from a low-mass companion ($\lesssim1\,\msun$) to a neutron star via Roche-lobe overflow, forming an accretion disc which feeds nuclear fuel to the neutron star surface. The base of the accreted layer is buried deeper under new material and heated to the point of thermonuclear runaway \citep{woosley_-ray_1976, joss_x-ray_1977}. The heat released by the rapid fusion of the accreted layer is observable as a burst of X-rays, lasting approximately 10--100 s. Fresh fuel is then accreted onto the ashes, and bursts recur within hours to days \citep[for reviews, refer to][]{lewin_x-ray_1993, strohmayer_new_2003, galloway_thermonuclear_2008}.

As each new layer of fuel is buried deeper, its hydrogen is steadily converted to helium via the beta-limited (hot) CNO cycle. If the burst recurrence time is longer than the time to deplete hydrogen, the burst will ignite in a deep helium layer \citep[Case 2, ][]{fujimoto_shell_1981}. This class of helium bursts reach Eddington luminosity (\sub{L}{Edd}) and exhibit photospheric radius expansion \citep[PRE;][]{tawara_very_1984,lewin_precursors_1984}. Their lightcurves feature rapid onsets ($\lesssim1\,\mathrm{s}$), broad plateaus ($\approx 10\,\mathrm{s}$), and short tails ($\lesssim 30\,\mathrm{s}$) due to the absence of extended \textsl{rp}-process burning.

Some X-ray burst systems, such as \gs{} \citep{ubertini_bursts_1999}, accrete and produce bursts at a consistent rate. X-ray burst modelling to date has focused on stable accretion rates \citep[e.g.,][]{woosley_models_2004, heger_models_2007, keek_multi-zone_2011}, and the dependence of burst properties on those rates \citep[][]{lampe_influence_2016}. Transient X-ray binaries, on the other hand, can remain dormant for years at a time and experience unstable surges of accretion known as outbursts. The addition of fresh material to the neutron star surface, as with stable accretors, can trigger series of X-ray bursts.

\saxj{} was the first accreting millisecond pulsar (AMXP) to be observed \citep{in_t_zand_discovery_1998, wijnands_millisecond_1998, chakrabarty_two-hour_1998}, and undergoes month-long outbursts every $2-3$ years \citep[e.g.,][]{wijnands_observational_2004, galloway_accretion-powered_2006, hartman_long-term_2008, hartman_luminosity_2009, patruno_accreting_2012, patruno_accelerated_2012, patruno_radio_2017}. During the well-sampled outburst of 2002 October, four thermonuclear X-ray bursts were observed \citep{chakrabarty_nuclear-powered_2003}. Subsequent modelling determined these to be helium bursts \citep[][hereafter G06]{galloway_helium-rich_2006}. The bursts from this event have been proposed as a standard test case for numerical modelling \citep[Case 2,][]{galloway_thermonuclear_2017}.

Using a semi-analytic model with a one-zone ignition criterion \citep[described in][]{cumming_rotational_2000}, \citetalias{galloway_helium-rich_2006} matched models to the observed burst properties and constrained the distance of the system to $3.5\pm 0.1\,\mathrm{kpc}$. In order to solve for the ignition conditions, the accretion rate was averaged between bursts. Because these models were computationally inexpensive, many thousands could efficiently explore parameter-space. Despite these advantages, the semi-analytic nature of the model lacks chemical and thermal inertia from one burst to the next.

To improve upon this modelling of the 2002 October outburst, we present a multi-zone simulation produced with the \kepler{} code, which tracks the composition and thermal state of the envelope as the outburst evolves \citep{woosley_models_2004}. This is the first application of time-dependent accretion rates to multi-zone burst simulations, allowing us to capture the thermal and chemical inertia of the envelope throughout an accretion episode.

In Section~\ref{sec:method}, we describe the X-ray data from the 2002 October outburst, the code we used to simulate the bursts, the system parameters chosen for the simulation, how we constructed the accretion rate curve, and how burst properties were extracted for comparison with observations. In Section~\ref{sec:results}, we examine the effect that a time-varying accretion rate has compared to a constant rate, and compare the simulated burst properties with those observed, including times-of-arrival, fluences, and lightcurves. We summarise our results in Section~\ref{sec:conclusion}, and discuss planned improvements to the model.

\begin{figure}
  \centering
	\includegraphics[width=\columnwidth]{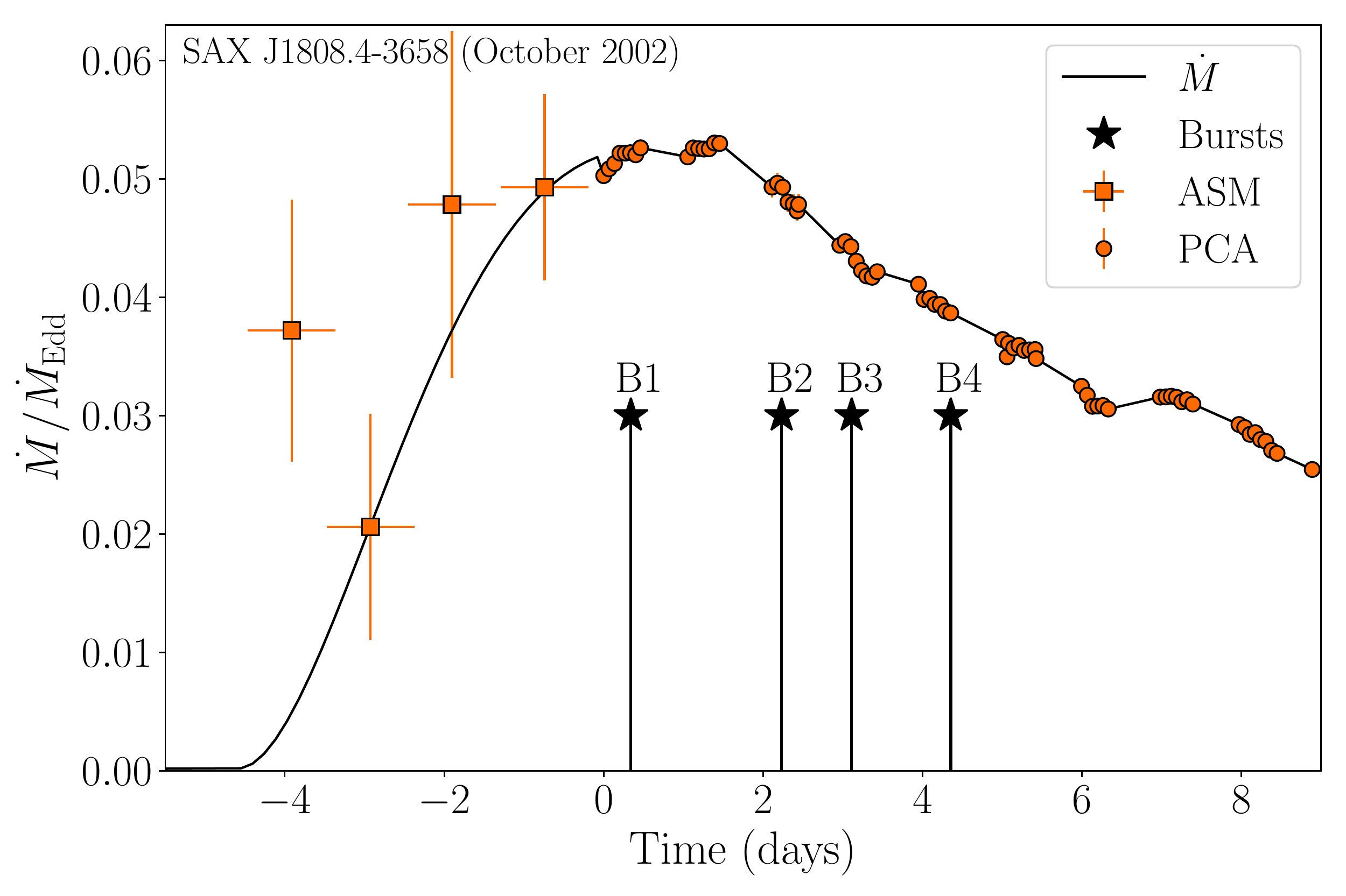}
	\caption{The accretion rate estimated from \textit{RXTE} observations during the 2002 October outburst, as a fraction of the Eddington-limited rate ($\sub{\mdot}{Edd}=1.75\ten{-8} \, \msun \, \mathrm{yr}^{-1}$), assuming $d=3.5\, \mathrm{kpc}$ and $\sub{\xi}{p}$. Time has been zeroed to the start of PCA observations at MJD 52562.07296, and shown are the times of the four observed bursts. Linear interpolation has been used within the high-resolution PCA data, while a toy curve has been inserted as a stand-in for the rise, which was not observed by the PCA. Note that while ASM data have been plotted for reference, the rise curve was not explicitly fit to them.}
	\label{fig:accrate}
\end{figure}


\section{Method}
\label{sec:method} 

\subsection{Observational data}
\label{sec:instruments}
We used data from the 2002 October outburst taken with the Proportional Counter Array \citep[PCA;][]{jahoda_-orbit_1996} and the All Sky Monitor \citep[ASM;][]{levine_first_1996} of the \textit{Rossi X-ray Timing Explorer} (\textit{RXTE}). The PCA instrument is composed of five proportional counter units (PCUs) sensitive to photon energies of $2-60\,\mathrm{keV}$. Two components of the PCA data were used: the persistent accretion flux, \sub{F}{p} (\S.~\ref{sec:accrate}), and the time-resolved burst lightcurves, \sub{F}{b} (\S.~\ref{sec:bprops}). The purpose of the ASM instrument is primarily to trigger alerts of transient events, and produces only low signal-to-noise data. It scans the sky every $90$-minute orbit, and scans a given object roughly $5-10$ times each day in $90$-second exposures, from which $1$-day average count rates are calculated.

The burst data were taken from the catalogue of \cite{galloway_thermonuclear_2008}, which was since re-analysed for the Multi INstrument Burst ARchive (MINBAR, in-development)\footnote{\url{http://burst.sci.monash.edu/minbar/}}. This re-analysis consisted of fitting absorbed blackbody models over the range $2.5-20 \,\mathrm{keV}$, with the neutral column density fixed at $n_\mathrm{H}=1.2\times10^{21} \pcm$ \citep{wang_optical_2001}.
Updated PCA response matrices (v11.7\footnote{\url{http://heasarc.gsfc.nasa.gov/docs/xte/pca/doc/rmf/pcarmf-11.7/}}) were used, and the recommended systematic error of $0.5$ percent was adopted.
Churazov weighting was employed to address the issue of low-count spectra in \textsc{xspec} \citep{dorman_redesign_2001}.
Deadtime was estimated using the Standard-1 mode data\footnote{following the recipe at \url{http://heasarc.gsfc.nasa.gov/docs/xte/recipes/pca\_deadtime.html}}, and the exposure time was then reduced by the deadtime correction factor, contributing an additional $25-30$ percent to the photon flux at the burst peaks.

For the persistent flux (\sub{F}{p}), $n_\mathrm{H}$ was fixed, and updated response matrices and deadtime correction used, as with the burst data. Additionally, spectra were averaged over each observation separately for each PCU, excluding the burst times. Each spectrum was then fitted with one of a family of models, including blackbody+powerlaw or Comptonisation, often adopting a Gaussian component to model Fe K$\alpha$ emission around $6.4 \,\mathrm{keV}$. The models were then integrated over the range 3-25 keV, and the bolometric flux estimated.


\subsection{Numerical method}
\label{sec:kepler}
To simulate the neutron star envelope during an accretion outburst, we used the 1D stellar hydrodynamics code \kepler{}, which models a grid of Lagrangian zones in the radial direction \citep{weaver_presupernova_1978,woosley_models_2004}. Each zone, representing a spherically symmetric shell of stellar material, has its own isotopic abundances and thermal properties. Convection of heat and nuclei between zones is modelled using mixing length theory, where a time-dependent diffusion coefficient is set by the convective velocity \citep[implementation described in][]{heger_presupernova_2000}.

\kepler{} uses an adaptive nuclear network that can track the reactions between more than $1,000$ isotopes up to the proton drip line \citep{rauscher_nucleosynthesis_2002}.  Isotopes are automatically added and removed from the network as needed \citep{woosley_models_2004}.  This allows us to efficiently model the $\beta$-limited CNO cycle, the $3\alpha$-process, the $\alpha$\textsl{p}-process, and the \textsl{rp}-process \citep{cyburt_jina_2010,cyburt_dependence_2016}.

\kepler{} also uses an adaptive spatial grid, in which zones are actively split or combined at each timestep in order to maintain resolution of thermodynamic gradients. Multiple criteria govern this rezoning; we impose a minimum zone thickness of 10 cm, a surface zone mass of $\sim 10^{18} \,\mathrm{g}$, and the above-mentioned accretion depth of $10^{19} \, \mathrm{g}$. These were chosen to avoid needlessly creating large numbers of zones, while maintaining consistency of the resulting burst properties. The model described in \S~\ref{sec:results} has 71 initial zones, growing as mass is accreted to 122 zones at the time of the first burst, and to 174 zones by the end of the simulation.

The simulation domain extends from the NS photosphere to the deep ocean near the crust, covering column depths of $10^{4} \lesssim y \lesssim 10^{12}\,\mathrm{g} \pcm$ ($10^4 \lesssim \rho \lesssim 10^9\,\mathrm{g\,cm^{-3}}$). Accretion-driven heating in the crust from electron captures and pycnonuclear reactions is included as a luminosity at the base of the grid, \sub{L}{crust}. To set up the initial thermal state, the envelope is relaxed until thermal equilibrium is met between \sub{L}{crust} and the luminosity at the surface. 

Nuclear reactions in the envelope are then switched on, and accretion is simulated by adding mass to the zone with an exterior mass coordinate of $10^{19} \, \mathrm{g}$ ($y \approx 8 \ten{5} \,\mathrm{g} \pcm$) at the rate $\mdot$, which may evolve with time. This depth is chosen to avoid unnecessary re-zoning at the surface. Above this, the accreted composition is advected, and heating from accretion and compression is included \citep{keek_multi-zone_2011}.

The accreted fuel is composed of $^1$H, $^4$He, and $^{14}$N, given by their mass fractions, \hyd, \hel, and \cno, respectively. We choose $^{14}$N for simplicity, because it is the most abundant CNO metal in solar material, and the hot-CNO cycle rapidly stabilises to equilibrium values of $^{14}$O and $^{15}$O.


\subsection{General relativistic corrections}
\label{sec:GR}
\kepler{} performs calculations in the local NS frame using Newtonian gravity. This approximation is acceptable because the grid spans only a thin surface shell of the neutron star, over which the gravitational acceleration, $g$, varies by $\lesssim 2\%$. However, in order to compare results with observations,  general relativity (GR) must be accounted for. Given a model Newtonian NS mass and radius ($M$, $R$), there are combinations of GR mass and radius ($M_\mathrm{GR}$, $R_\mathrm{GR}$) such that the surface gravity is equal under both regimes \citep[refer to Appendix B of][]{keek_multi-zone_2011}. In other words, the model can be considered equivalent to a neutron star with an `actual' mass and radius of $M_\mathrm{GR}$ and $R_\mathrm{GR}$. This is satisfied when
\begin{equation}
\label{eq:grav}
\frac{\mathrm{G} M}{R^2} = \frac{\mathrm{G} M_\mathrm{GR}}{R_\mathrm{GR}^2} (1+z),
\end{equation}
where the gravitational redshift factor is
\begin{equation}
1+z=\frac{1}{\sqrt{1-2\mathrm{G}M_\mathrm{GR}/(c^2 R_\mathrm{GR})}}.
\end{equation}
If we choose $M_\mathrm{GR}=M$, the solutions are simplified, and the luminosity and time can be converted from the model's Newtonian frame ($L$, $t$) to a distant observer frame ($L_\infty$,~$t_\infty$) with
\begin{equation}
\label{eq:gr}
L_\infty = \frac{L}{1+z}, \qquad t_\infty = t\,\left(1+z\right).
\end{equation}


\begin{figure}
  \centering
	\includegraphics[width=\columnwidth]{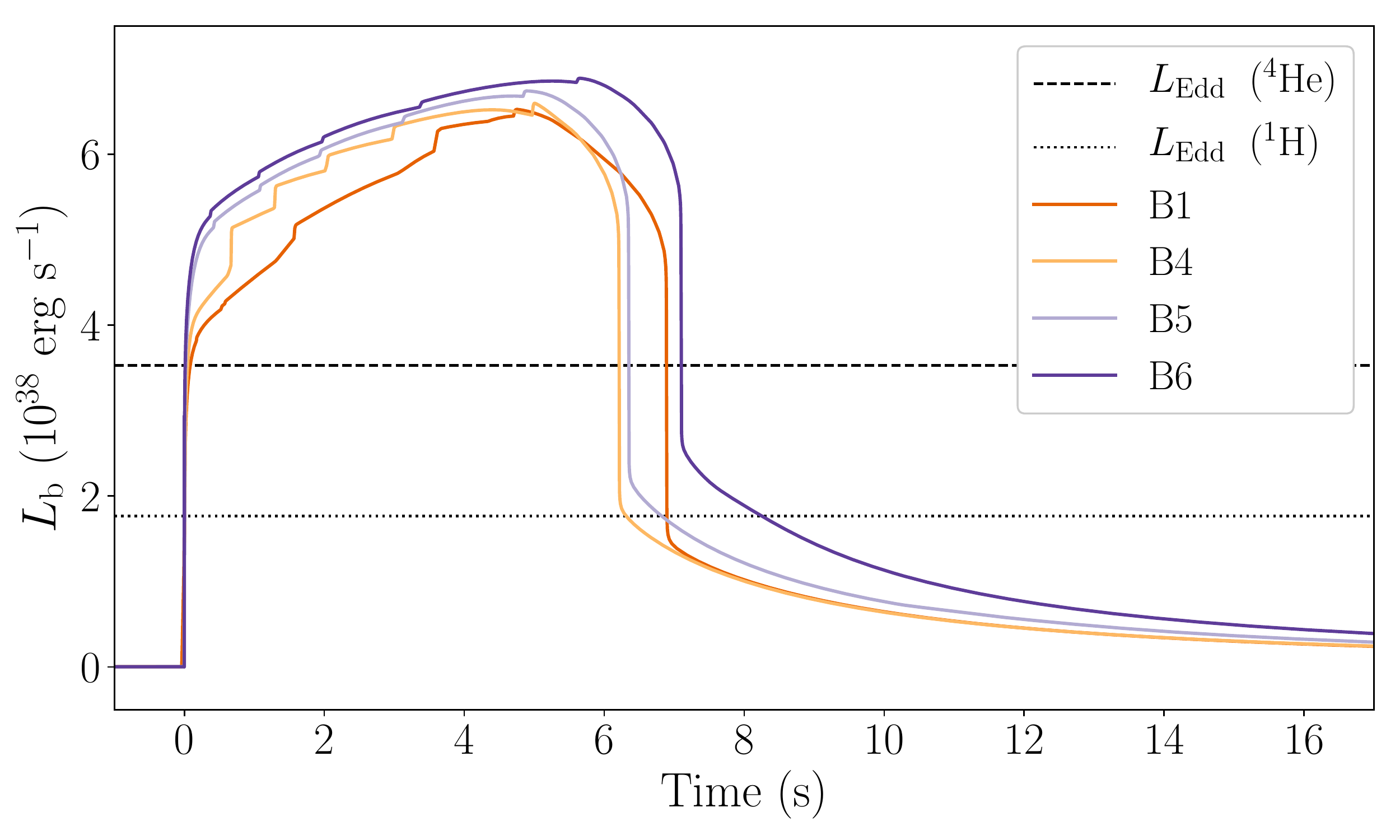}
	\caption{The raw simulated burst lightcurves, in the frame of the model. PRE bursts often exceed the Eddington luminosity in \kepler{}, presumably because the physics of photospheric expansion and contraction are not accurately captured by the simple atmosphere, in addition to the lack of an outflow/wind mechanism. Low surface-resolution causes the visible steps during the peak, due to convection switching on as it spreads through these outer zones. To compare with the observations, we manually truncated the lightcurves at $\sub{L}{Edd}\, (^4\mathrm{He})$ during analysis (\S~\ref{sec:analysis}; Fig.~\ref{fig:lightcurves}).}
	\label{fig:superedd}
\end{figure}

\begin{table}
    {\centering
    \caption{Summary of parameters used in the model. Please note that these should not yet be considered best fit values to the system.}
    \label{tab:params}
    \begin{tabular}{l|l|l|l}
        \hline
        Quantity        &                 & Units & Description\\
        \hline
         $g$            &  1.86           & $10^{14}\, \mathrm{cm\, s^{-2}}$ & Surface gravity \\
         $M$            &  1.4$^\ast$     & $\msun$                          & NS mass \\
         $R$            &  11.2$^\ast$    & km                               & NS radius \\
         $\sub{Q}{b}$   &  0.3            & $\mathrm{MeV\,nucleon^{-1}}$     & Crustal heating \\
         $\hyd$         &  0.44           & Mass fraction                    & Accreted hydrogen \\
         $\cno$         &  0.02           & Mass fraction                    & Accreted metallicity \\
         $d$            &  3.5$^\dagger$  & kpc                              & Distance \\
         $\sub{\xi}{p}$ &  1.1$^\dagger$  & --                               & Persistent anisotropy \\
         \hline
    \end{tabular}\\
    }
    $\ast$ Effective GR-corrected values. Other combinations are still valid if they preserve $g$, but will alter (1+z) and the conversion of model results to observable values (\S~\ref{sec:GR}) \\
    $\dagger$ Assumed values for inferring the accretion rate from persistent flux. Other combinations are still valid (\S~\ref{sec:accrate})\\
\end{table}

\subsection{Model parameters}
\label{sec:model}
The model parameters are summarised in Table~\ref{tab:params}. 

We adopt a gravitational mass of $M=M_\mathrm{GR}=1.4$ $\msun$ and a Newtonian model radius of $R=10$ km, equivalent to $R_\mathrm{GR} \approx 11.2$ km (Equation~\ref{eq:grav}). This gives a surface gravity of $g \approx 1.858\ten{14}\,\mathrm{cm}\,\mathrm{s}^{-2}$ and redshift of $1+z \approx 1.259$. Note that the model can still be considered equivalent to any other pair of \sub{M}{GR} and \sub{R}{GR} that satisfy Equation~\ref{eq:grav}.

To set up the initial envelope, the inner portion is composed of an iron substrate between column depths of $10^{8} \lesssim y \lesssim 10^{12}\,\mathrm{g\pcm}$ ($10^6 \lesssim \rho \lesssim 10^9\,\mathrm{g\,cm^{-3}}$). Nuclear reactions are not calculated in the substrate, and it primarily acts as a thermal sink during bursts, representing the ocean of prior burst ashes. Above the substrate, between $10^{4} \lesssim y \lesssim 10^{8}\,\mathrm{g\pcm}$ ($10^4 \lesssim \rho \lesssim 10^6\,\mathrm{g\,cm^{-3}}$), we add $^4$He to represent leftover fuel from the tail of the previous outburst. The size of this fuel layer could be varied in future studies.

Crustal heating from the inner boundary evolves with accretion rate according to $\sub{L}{crust}=Q_\mathrm{b}\mdot$, where $Q_\mathrm{b}$ is the specific energy yield from reactions in the crust. We adopt a value of $Q_\mathrm{b}=0.3\,\mathrm{MeV\,nucleon^{-1}}$, following the best-fitting model from \citetalias{galloway_helium-rich_2006}. To initialise the envelope, we use this value with the long-term average accretion rate of $10^{-11}\, \msun\, \mathrm{yr^{-1}}$ (\citetalias{galloway_helium-rich_2006}), until the layer is in thermal equilibrium.

The observed burst energetics indicate an average hydrogen composition of $\langle X \rangle \approx 0.1$ at the time of ignition (\citetalias{galloway_helium-rich_2006}). The initial composition of the accreted fuel, however, is less well-constrained because there is a degeneracy between the $\hyd$ and $\cno$ which result in the above $\langle X \rangle$ due to stable hot-CNO burning. For the purposes of this study, we choose a metallicity of $\cno=0.02$, following the best-fitting model from \citetalias{galloway_helium-rich_2006}. Initial tests of \hyd{} in the range of $0.49-0.62$ (from the best-fitting $1\sigma$ range of  \citetalias{galloway_helium-rich_2006}) indicated that our models required a lower value to reproduce the observed burst timings. The model presented here has $\hyd=0.44$, with the remaining composition being $\hel=0.54$.

Accretion discs can cause anisotropies by scattering and blocking X-ray emission from the NS surface, changing the apparent luminosity to an observer \citep{fujimoto_angular_1988, he_anisotropy_2016}. The strength of the anisotropy is dependent on $i$, the inclination of the binary system to the observer's line-of-sight. The inclination is typically poorly constrained in LMXBs, with the value for \saxj{} inferred to be $50^\circ\lesssim i \lesssim 80^\circ$ \citep{chakrabarty_two-hour_1998, wang_optical_2001, bildsten_brown_2001, ibragimov_accreting_2009}. The effect is represented as a scaling factor, $\xi$, which we include in the conversion between the model luminosity and the observed flux, given by 
\begin{equation}
\label{eq:flum}
L=4\mathrm{\pi} d^2 \xi F.    
\end{equation}
 Independent factors are used for the persistent emission (\sub{\xi}{p}) and the burst emission (\sub{\xi}{b}), because the emitting region is not necessarily the same for both mechanisms. We used \sub{\xi}{p}{} when calculating \mdot{} from persistent flux, and \sub{\xi}{b}{} when calculating the burst flux from the model burst luminosity.


\begin{figure}
  \centering
	\includegraphics[width=\columnwidth]{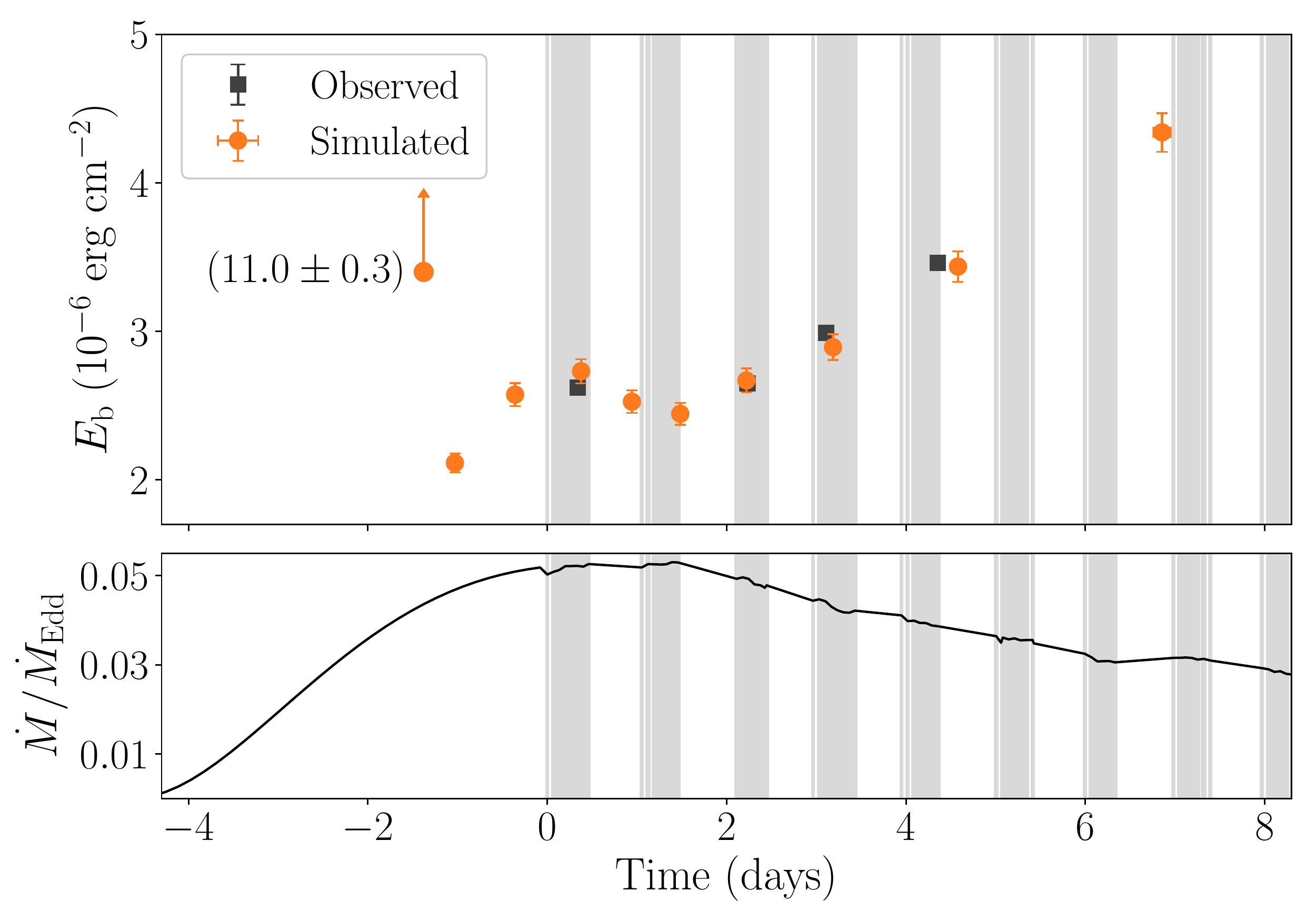}
	\caption{Upper panel: Fluence, $\Eb$, of the modelled burst sequence against the four observed bursts. Lower panel: the time-varying accretion rate over the event, as a fraction of the Eddington-limited accretion rate. The vertical grey bands indicate when the telescope was collecting data. Note that extra bursts predicted by the model fall outside these observing windows. As is typical for \kepler{} models, the first burst was anomalously energetic, and its off-axis $\Eb$ is indicated next to the arrow. The fluences have been calculated from the burst energy with the scaling factor $\mathrm{c_2} = 4 \pi d^2 \sub{\xi}{b} \approx 1.05\ten{45}\, \mathrm{cm^2}$, chosen such that the RMS error with observations is minimised (\S~\ref{sec:bprops}).}
	\label{fig:main}
\end{figure}

\begin{figure}
  \centering
	\includegraphics[width=\columnwidth]{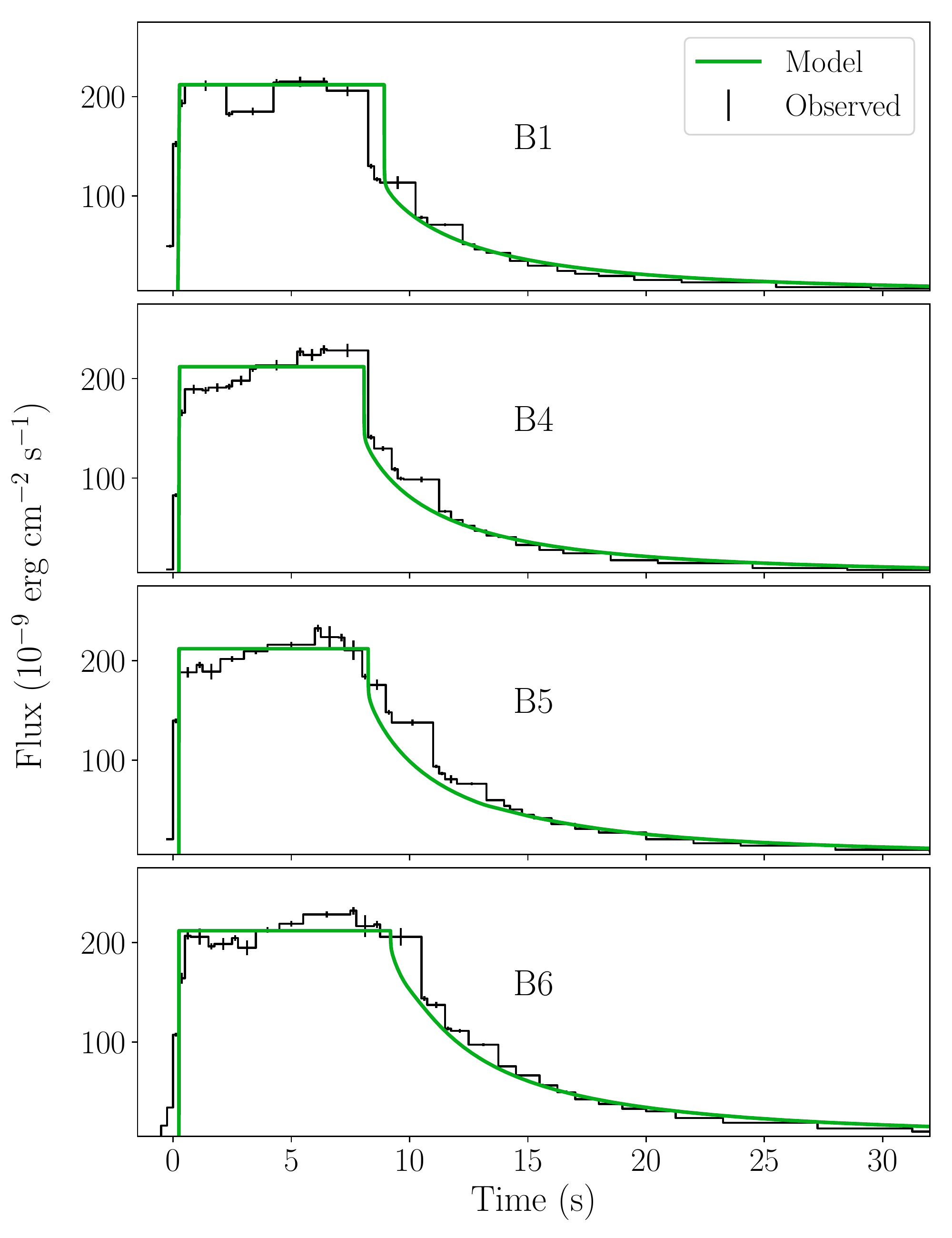}
	\caption{The lightcurves for each modelled burst (solid green curve) against its observed counterpart (black bins). The model luminosity has been converted to an observed flux using the scaling factor $\mathrm{c_2}=4\pi d^2 \sub{\xi}{b} = 1.05\ten{45}\, \mathrm{cm^2}$, chosen such that the burst fluences best match the observations (\S~\ref{sec:bprops}).}
	\label{fig:lightcurves}
\end{figure}


\subsection{Accretion history}
\label{sec:accrate}
We inferred the accretion history of the 2002 October outburst from the observed persistent flux, \sub{F}{p}. Accretion onto the surface generates a luminosity of
\begin{equation}
\label{eq:lacc}
  L_\mathrm{acc}=-\mdot\phi \quad \lunits,  
\end{equation}
where $\phi=-\mathrm{c}^2z/(1+z)\approx -0.2\,\mathrm{c}^2$ is the gravitational potential at the NS surface. Using Equations~\ref{eq:gr}~and~\ref{eq:flum}, the accretion rate in the model frame is then given by
\begin{equation}
\label{eq:mdot}
  \mdot = \frac{-4 \mathrm{\pi} d^2 \sub{\xi}{p} (1+z)}{\phi} \sub{F}{p}.
\end{equation}
We can rewrite this expression in terms of a conversion constant,
\begin{equation}
\label{eq:mdotscaled}
  \mdot = \frac{-\mathrm{c_1} (1+z)}{\phi} \sub{F}{p},\quad
  \mathrm{c_1} = 4 \mathrm{\pi} d^2 \sub{\xi}{p}.
\end{equation}

For this model, we chose $d=3.5\,\mathrm{kpc}$ (\citetalias{galloway_helium-rich_2006}), and $\sub{\xi}{p} = 1.1$, which is the predicted anisotropy factor for an inclination of $55^\circ \lesssim i \lesssim 60^\circ$ \citep[Fig. 8 of][]{he_anisotropy_2016}. Thus, we have a conversion constant of $\mathrm{c_1} \approx 1.612\ten{45} \, \mathrm{cm}^2$. Note that this model is still applicable to other combinations of $d$ and $\sub{\xi}{p}$ that preserve $\mathrm{c_1}$.

PCA observations did not commence until the peak of the outburst, and so the precise onset of accretion is ambiguous. For reference, the rise of the subsequent 2005 June outburst was observed to last 5 days \citep{hartman_long-term_2008}. For the 2002 October outburst, only 1-day average count rates from the ASM are available, constraining the rise length to $4-5$ days. With these considerations, we substituted a toy curve for the accretion rise with a length of approximately 4 days. We expect that differences in the chosen onset should primarily influence the first burst or two, which likely went unobserved (\S~\ref{sec:bprops}). Nevertheless, we plan to investigate the effect of rise length and shape in a future study.

Combined with the PCA data and Equation~\ref{eq:mdot}, we thus obtained a continuous $\mdot(t)$ curve for input to the model (Fig.~\ref{fig:accrate}).


\subsection{Burst properties}
\label{sec:analysis}
We ran \kepler{} with the above inputs, and obtained a sequence of bursts over the course of the outburst. We then extracted the burst lightcurves from the modelled NS surface luminosity, and calculated their properties in a process similar to \citet{lampe_influence_2016}. 

The recurrence time, \dt, is the time from one burst to the next. The burst energy, $\sub{E}{nuc}$, is obtained by integrating over the lightcurve. This translates to the observed burst fluence, $\Eb$, via
\begin{equation}
  \Eb = \frac{1}{4\pi d^2 \sub{\xi}{b}}\sub{E}{nuc}\,.
\end{equation}
Similar to Equation~\ref{eq:mdotscaled}, this expression can also be written in terms of a scaling factor,
\begin{equation}
\label{eq:eb}
  \Eb = \frac{1}{\mathrm{c_2}} \sub{E}{nuc} \;\;\text{with}\;\; \mathrm{c_2} = 4\pi d^2 \sub{\xi}{b}.
\end{equation}

The profiles of PRE burst lightcurves in \kepler{} noticeably deviate from observations, likely due to the simple atmosphere \citep[see Model \textit{Zm} in][]{woosley_models_2004}. In our model, the surface luminosity exceeded the Eddington limit by up to a factor of two, followed by a steep drop (Fig.~\ref{fig:superedd}). Luminosity in excess of Eddington should drive further radius expansion, or even be converted into a wind, a mechanism which these models lack. In order to compare our results with observations, we manually truncated the lightcurves at the Eddington luminosity for pure helium, $\sub{L}{Edd}=3.53\ten{38} \lunits$, the inferred limit reached for PRE bursts \citep{kuulkers_photospheric_2003}.


\section{Results}
\label{sec:results}
We present here a model which closely reproduced the observed burst times. We must emphasise that we present this not as a best-fitting model to the observations, but to demonstrate the feasibility of modelling bursts under varying accretion rates. A more detailed and systematic exploration of model parameters is planned as a future study.

The model produced a sequence of ten X-ray bursts over the course of the outburst (Table \ref{tab:main}, Fig.~\ref{fig:main}). The bursts have been assigned labels to aid in discussion. Seven are labelled sequentially from B1 to B7, where B1 is the closest in time to the first observed burst. The observed bursts are labelled O1, O4, O5, and O6, to correspond with their closest model bursts. Three bursts occur prior to B1, and are labelled P1, P2, and P3, in reverse order from B1, in anticipation of future simulations which may produce more or fewer such bursts.

We have applied GR corrections to the modelled burst properties, such that they correspond to a distant observer (\S~\ref{sec:GR}). Error bars for the observational data are $1\sigma$ uncertainties. We have set the model uncertainties to 3 percent, which is the typical $1\sigma$ variation in modelled burst trains when all input parameters (including $\mdot$) are held constant (\S~\ref{sec:avg}). The uncertainties in the modelled burst arrival times were obtained by propagating the 3 percent uncertainty in \dt{} along the burst train (the uncertainty for the first burst was simply taken to be that of the following recurrence interval). Additional model uncertainties due to the observational uncertainties in distance, inclination etc. are not considered for the purpose of this paper, because we are not yet attempting parameter estimation.


\subsection{Varying versus averaged accretion rates}
\label{sec:avg}
\begin{figure}
  \centering
	\includegraphics[width=\columnwidth]{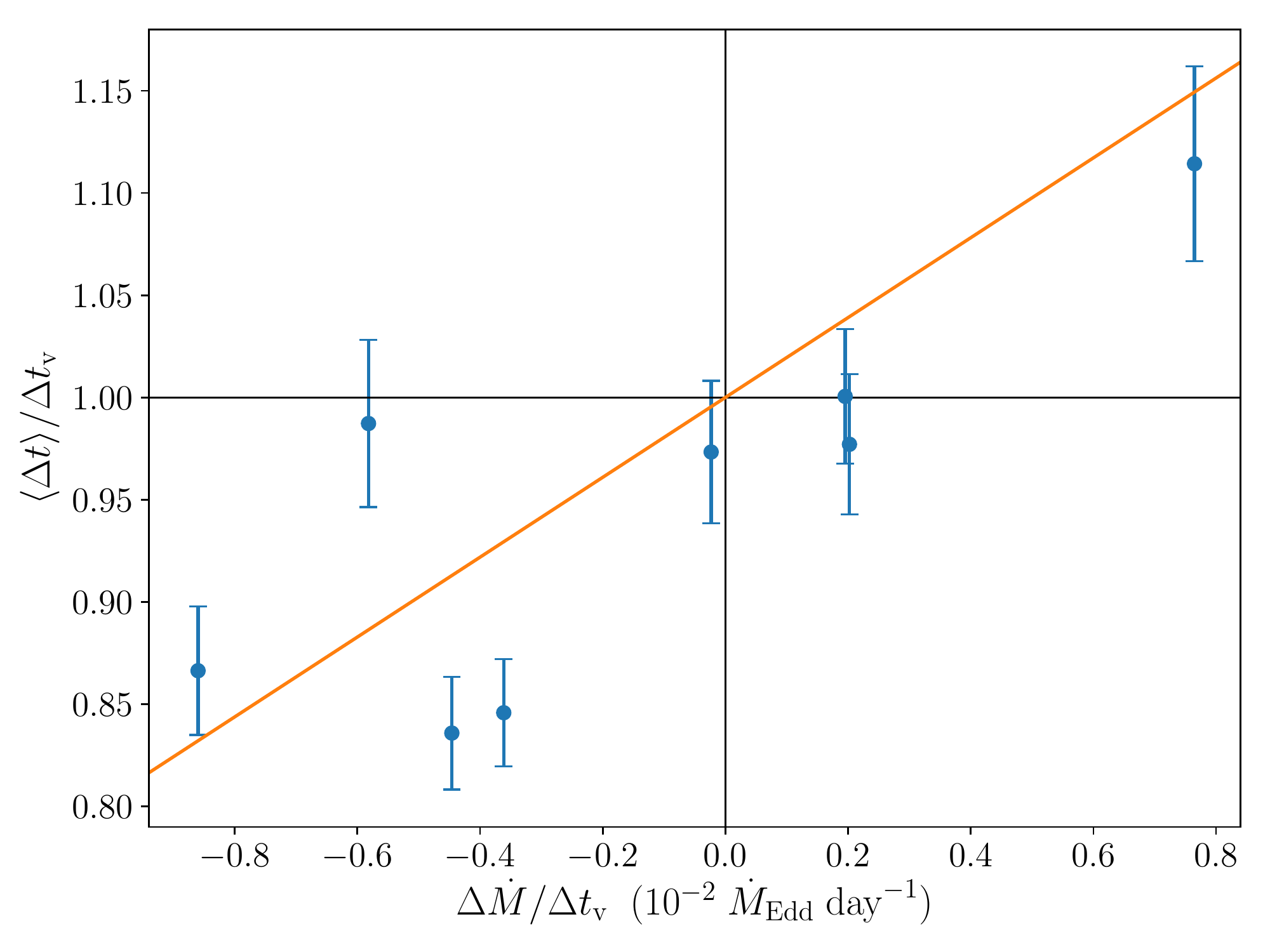}
	\caption{The result of using a constant \mdotavg{} in place of a varying $\mdot(t)$. The vertical axis is the ratio of the recurrence time from a constant accretion rate (\dtavg) to the recurrence time from a varying accretion rate ($\Delta \sub{t}{v}$). On the horizontal axis is the average slope of $\mdot(t)$ for the interval. The trendline is a weighted least-squares regression that has been forced through the point $(0,1)$, and has a slope of $0.195$.}
	\label{fig:ratio}
\end{figure}

We performed a comparison test in order to check for any difference in results between a varying and an averaged accretion rate. For each burst interval in the model, $\sub{\Delta t}{v}$, we calculated the average accretion rate, \mdotavg. We then independently restarted the simulation at the beginning of each interval, with \mdot{} now fixed at \mdotavg. Once a sequence of $10-15$ bursts were produced at each \mdotavg{}, we calculated the mean recurrence times, \dtavg. We then compared these with the original recurrence times via the ratio $\dtavg / \sub{\Delta t}{v}$, and calculated the average slope of $\mdot(t)$ for each interval, given by $\Delta \mdot / \sub{\Delta t}{v}$ (Fig.~\ref{fig:ratio}). The standard deviation in burst properties for each \dtavg{} was typically $\approx 3$ percent, and we have adopted this as the standard model uncertainty. We have excluded the first burst interval (P3-P2) from this comparison because it is abnormally short, with a recurrence time almost half the length of the next, despite a lower accretion rate. We attribute this to extra heating from the very large preceding burst, P3.

The variables $\dtavg / \sub{\Delta t}{v}$ and $\Delta \mdot / \sub{\Delta t}{v}$ have a Spearman's rank correlation coefficient of $r_s = 0.60$, with a p-value of $p=0.12$. A least-squares linear regression, weighted with the uncertainties and forced through the point $(0,1)$, returns a slope of 0.195. The correlation appears to have a low significance, although it is somewhat inconclusive with such a small sample. The scatter may be due in part to the small-scale variations in $\mdot(t)$ within each interval, given that $\Delta \mdot / \sub{\Delta t}{v}$ is itself an approximation to the slope. Nevertheless, this tentatively suggests that using a constant \mdotavg{} may systematically overestimate \sub{\Delta t}{v} when $\mdot(t)$ is increasing, and underestimate \sub{\Delta t}{v} when $\mdot(t)$ is decreasing. This discrepancy could have a significant effect on predictions, because burst properties are strongly dependent on the recurrence time. It's possible that the relationship is dependent on other parameters such as $\sub{Q}{b}$, \hyd{} and $\cno$. With further study, a correction factor could account for this systematic discrepancy when an averaged accretion rate can't be avoided, as with the semi-analytic models of \citetalias{galloway_helium-rich_2006}.

\begin{table}
  {\centering
  \caption{Burst properties for the model with $\hyd=0.44$ (for a description of burst labels, see \S~\ref{sec:results}). Observed bursts and their modelled counterparts are highlighted in bold.}
  \label{tab:main}
  \begin{tabular}{lccc}
    \hline
    \hline
    Burst &	$t$ & $\Delta t$ & $\Eb$\\
       & (h) & (h)    & ($10^{-6} \flunits$)   \\
    \hline
    P3  & $-33.1 \pm 0.3$ & ---      & $11.0 \pm 0.3$ \\
    P2  & $-24.7 \pm 0.3$ & $8.4 \pm 0.3$ & $2.11 \pm 0.06$ \\
    P1  & $-8.6 \pm 0.5$ & $16.1 \pm 0.5$ & $2.57 \pm 0.08$ \\
    \noalign{\medskip}
    {\bf B1} & $9.0 \pm 0.8$  & $17.6 \pm 0.5$  & $2.73 \pm 0.08$ \\
    {\bf O1} & 8.2$^\ast$ & ---       & $2.620 \pm 0.021$ \\
    \noalign{\medskip}
    B2  & $22.7 \pm 0.9$ & $13.6 \pm 0.4$ & $2.53 \pm 0.08$ \\
    B3  & $35.6 \pm 0.9$ & $12.9 \pm 0.4$ & $2.45 \pm 0.07$ \\
    \noalign{\medskip}
    {\bf B4} & $53.3 \pm 1.1$  & $44.3 \pm 0.8^\dagger$ & $2.67 \pm 0.08$ \\
    {\bf O4} & 53.6       & $45.4$      & $2.649 \pm 0.018$ \\
    \noalign{\medskip}
    {\bf B5} & $76.5 \pm 1.3$  & $23.2 \pm 0.7$  & $2.89 \pm 0.09$ \\
    {\bf O5} & 74.7       & 21.1       & $2.990 \pm 0.017$ \\
    \noalign{\medskip}
    {\bf B6} & $109.9 \pm 1.6$ & $33.4 \pm 1.0$  & $3.44 \pm 0.10$ \\
    {\bf O6} & 104.5      & 29.8       & $3.460 \pm 0.022$ \\
    \noalign{\medskip}
    B7    & $165 \pm 2$   & $54.6 \pm 1.6$ & $4.34 \pm 0.13$ \\  
    \hline
  \end{tabular}
  }
  \\
  $^\ast$Observational time uncertainties are $< 1$ s and are excluded for clarity\\
  $^\dagger$Total interval between B1--B4, to compare with the observed value.\\
\end{table}


\subsection{Comparison with observed bursts}
\label{sec:bprops}
The four observed burst times are matched to within $0.85\,\mathrm{h}$, $0.27\,\mathrm{h}$, $1.8\,\mathrm{h}$, and $5.4\,\mathrm{h}$, respectively. The RMS error for the four bursts is $2.88\,\mathrm{h}$, in comparison to their recurrence times of $18 \lesssim \Delta t \lesssim 33\,\mathrm{h}$. An additional source of discrepancy here may be the gaps in PCA data. Because we have used linear interpolation in the PCA gaps, any unobserved variation in \mdot{} is not captured in the model.

Six extra bursts were predicted in addition to the four observed. Three of these (P1 -- P3) were during the accretion rise, preceding O1. Two intervening bursts (B2, B3) fell between O1 and O4, and a final burst (B7) fell during the outburst tail, $\approx 2.5\,\mathrm{days}$ after O6. All extra bursts fell within gaps in the PCA data, and so can't immediately be ruled out by observations. The two intervening bursts, B2 and B3, agree with the predictions of \citetalias{galloway_helium-rich_2006}, who concluded that two bursts likely occurred between O1 and O4. Our model also suggests that several other bursts may have been missed.

When converted from $\sub{E}{nuc}$ via Equation~\ref{eq:eb}, the burst fluences, $\Eb$, have a minimised RMS error with the observations when $\mathrm{c_2} \approx 1.05 \ten{45}\,\mathrm{cm^{2}}$. Following this scaling, the RMS error of the four bursts is $0.11\ten{-6} \flunits$, in comparison to their fluence range of $2.7 \lesssim \sub{E}{b} \lesssim 3.4 \flunits$. Overall, the model reproduces the observed trend of increasing $\Eb$ with recurrence time (Fig.~\ref{fig:main}). This is consistent with larger fuel layers accumulating due to a lowering accretion rate and a cooling envelope. 

From Equation~\ref{eq:mdotscaled} and Equation~\ref{eq:eb}, we have
\begin{equation}
  \label{eq:xi}
  \frac{\sub{\xi}{p}}{\sub{\xi}{b}} = \frac{\mathrm{c_1}}{\mathrm{c_2}} ,
\end{equation}
\begin{equation}
  \label{eq:distance}
  d = \sqrt{\frac{\mathrm{c_1}}{4 \pi \sub{\xi}{p}}} = \sqrt{\frac{\mathrm{c_2}}{4 \pi \sub{\xi}{b}}},
\end{equation}
where for this model $\mathrm{c_1}/\mathrm{c_2} \approx 1.536$. According to the thin accretion disc model from \citet[Model~a]{he_anisotropy_2016}, this anisotropy ratio occurs at an inclination of $i=67.7^\circ$, with $\sub{\xi}{p}=1.85$ and $\sub{\xi}{b}=1.21$. Using Equation~\ref{eq:distance}, this corresponds to a distance of $d=2.7 \,\mathrm{kpc}$, which is outside the range proposed by \citetalias{galloway_helium-rich_2006} of $3.5 \pm 0.1 \, \mathrm{kpc}$. This may be due to the inherent differences between the models, that the anisotropy was not considered in their study, or because our model is not yet a global best fit to the system. Additionally, a thin accretion disc may not be realistic for a transiently accreting LMXB, and other disc geometries produce different relationships between $\sub{\xi}{p}$ and $\sub{\xi}{b}$. For instance, a concave disc \citep[Model~d of][]{he_anisotropy_2016} predicts the same anisotropy ratio at $i=62.6^\circ$, with $\sub{\xi}{p}=1.34$ and $\sub{\xi}{b}=1.10$, corresponding to a distance of $d=3.1 \,\mathrm{kpc}$, closer to the accepted value.

Following Eddington-truncation and scaling, the burst lightcurves broadly reproduce the observed profiles, featuring sharp rise times and gradual decays (Fig.~\ref{fig:lightcurves}). The plateau peak has been fixed at $\sub{L}{Edd}$, and after scaling $\mathrm{c_2}$ to match the observed $\Eb$, it approximately agrees with the observed peaks.


\section{Conclusion}
\label{sec:conclusion}
We performed simulations of a neutron star envelope during an accretion outburst, using the time-dependent $\mdot(t)$ inferred from the 2002 October event of \saxj. A composition of $\cno=0.02$ and $\hyd=0.44$ reproduced the four observed burst arrival times with a RMS error of $2.88\,\mathrm{h}$, with recurrence times of $18 \lesssim \Delta t \lesssim 33\,\mathrm{h}$. The modelled sequence contained ten bursts, six of which did not correspond to observed bursts: three during the accretion rise; two between the first and second observed bursts; and one during the tail, $\approx 2.5$ days after the fourth observed burst. These extra bursts fell during times when the source was not being observed by \textit{RXTE}, which had a duty cycle of 38 percent for the outburst.

Due to the limitations of \kepler{} when simulating PRE bursts, the model-predicted luminosities were manually truncated at $\sub{L}{Edd}$ in order to compare other lightcurve features to observations. The lightcurves have rapid rise times ($<1\,\mathrm{s}$) and fast decays ($\approx 20 \,\mathrm{s}$), in agreement with the observed characteristics of helium-rich bursts. The burst fluences were reproduced with a RMS error of $0.11\ten{-6} \flunits$ after scaling by $\mathrm{c_2} = 4 \pi d^2 \sub{\xi}{b} \approx 1.05\ten{45} \mathrm{\, cm^2}$. 

To obtain similar recurrence times to the best-fitting models of \citetalias{galloway_helium-rich_2006}, we required a lower hydrogen fraction of $\hyd=0.44$, in comparison to $\hyd=0.54$. A difference is perhaps unsurprising given the degrees of complexity between the models. For example, the semi-analytic model of \citetalias{galloway_helium-rich_2006} solves for thermonuclear stability with a one-zone approximation, does not evolve bursts with time, and uses a simple expression for energy yield to calculate the total burst energy. Furthermore, our results suggest that using averaged accretion rates may overestimate the recurrence times for an increasing \mdot, and underestimate recurrence times for a decreasing \mdot. Efforts are currently underway to improve the semi-analytic model of \citetalias{galloway_helium-rich_2006} and its application to the 2002 October outburst. 

As mentioned, the model presented here is not yet a global fit to the data, and so posterior constraints on the system parameters are not yet possible. In a future study, a more systematic matching could be performed by varying $\cno$, $\sub{Q}{b}$, $M$, $R$, $\mathrm{c_1}$, $\mathrm{c_2}$, and the assumed accretion onset. The strength of the crustal heating, $\sub{Q}{b}$, may itself evolve with \mdot. In addition to the burst times, other properties such as the peak luminosities, lightcurve profiles, and $\Eb$ could be incorporated into the fitting routine to obtain a global likelihood value.

Matching the properties of multiple bursts over a single accretion event provides a new test bed for multi-zone models. Our simulations are the first to adopt a time-dependent $\mdot$, and demonstrate that existing burst models can be extended to transient accretion regimes. This may help to further constrain LMXB properties, including the NS mass and radius, the strength of crustal heating, and the distance and inclination of the system. Furthermore, constraining the fuel composition provides information about the composition of the companion star, and thus the evolutionary history of the binary. This could improve our understanding of the \saxj{} system, and more generally, the origin of accreting millisecond pulsars.

\section*{Acknowledgements}
This work was supported in part by the National Science Foundation under Grant No.\ PHY-1430152 (JINA Center for the Evolution of the Elements). This research was supported by an Australian Government Research Training Program (RTP) Scholarship. This paper utilises preliminary analysis results from the Multi-INstrument Burst ARchive (MINBAR), which is supported under the Australian Academy of Science's Scientific Visits to Europe program, and the Australian Research Council's Discovery Projects and Future Fellowship funding schemes. This research was supported in part by the Monash eResearch Centre and eSolutions-Research Support Services through the use of the MonARCH HPC Cluster. We thank Jordan He and Laurens Keek for providing data tables from their disc anisotropy models described in \citet{he_anisotropy_2016}, and Andrew Cumming and Adelle Goodwin for discussions about the analytic model used in \citetalias{galloway_helium-rich_2006}. AH was supported by an ARC Future Fellowship (FT120100363). ZJ would like to thank Michigan State University for their generous hospitality on a research visit, during which this manuscript was partially prepared.



\bibliographystyle{mnras}
\bibliography{MyLibrary} 

\begin{thebibliography}{}
\makeatletter
\relax
\def\mn@urlcharsother{\let\do\@makeother \do\$\do\&\do\#\do\^\do\_\do\%\do\~}
\def\mn@doi{\begingroup\mn@urlcharsother \@ifnextchar [ {\mn@doi@}
  {\mn@doi@[]}}
\def\mn@doi@[#1]#2{\def\@tempa{#1}\ifx\@tempa\@empty \href
  {http://dx.doi.org/#2} {doi:#2}\else \href {http://dx.doi.org/#2} {#1}\fi
  \endgroup}
\def\mn@eprint#1#2{\mn@eprint@#1:#2::\@nil}
\def\mn@eprint@arXiv#1{\href {http://arxiv.org/abs/#1} {{\tt arXiv:#1}}}
\def\mn@eprint@dblp#1{\href {http://dblp.uni-trier.de/rec/bibtex/#1.xml}
  {dblp:#1}}
\def\mn@eprint@#1:#2:#3:#4\@nil{\def\@tempa {#1}\def\@tempb {#2}\def\@tempc
  {#3}\ifx \@tempc \@empty \let \@tempc \@tempb \let \@tempb \@tempa \fi \ifx
  \@tempb \@empty \def\@tempb {arXiv}\fi \@ifundefined
  {mn@eprint@\@tempb}{\@tempb:\@tempc}{\expandafter \expandafter \csname
  mn@eprint@\@tempb\endcsname \expandafter{\@tempc}}}

\bibitem[\protect\citeauthoryear{Belian, Conner  \& Evans}{Belian
  et~al.}{1976}]{belian_discovery_1976}
Belian R.~D.,  Conner J.~P.,   Evans W.~D.,  1976, \mn@doi [The Astrophysical
  Journal Letters] {10.1086/182151}, 206, L135

\bibitem[\protect\citeauthoryear{Bildsten \& Chakrabarty}{Bildsten \&
  Chakrabarty}{2001}]{bildsten_brown_2001}
Bildsten L.,  Chakrabarty D.,  2001, \mn@doi [The Astrophysical Journal]
  {10.1086/321633}, 557, 292

\bibitem[\protect\citeauthoryear{Chakrabarty \& Morgan}{Chakrabarty \&
  Morgan}{1998}]{chakrabarty_two-hour_1998}
Chakrabarty D.,  Morgan E.~H.,  1998, \mn@doi [Nature] {10.1038/28561}, 394,
  346

\bibitem[\protect\citeauthoryear{Chakrabarty, Morgan, Muno, Galloway, Wijnands,
  van~der Klis  \& Markwardt}{Chakrabarty
  et~al.}{2003}]{chakrabarty_nuclear-powered_2003}
Chakrabarty D.,  Morgan E.~H.,  Muno M.~P.,  Galloway D.~K.,  Wijnands R.,
  van~der Klis M.,   Markwardt C.~B.,  2003, \mn@doi [Nature]
  {10.1038/nature01732}, 424, 42

\bibitem[\protect\citeauthoryear{Cumming \& Bildsten}{Cumming \&
  Bildsten}{2000}]{cumming_rotational_2000}
Cumming A.,  Bildsten L.,  2000, \mn@doi [The Astrophysical Journal]
  {10.1086/317191}, 544, 453

\bibitem[\protect\citeauthoryear{Cyburt et~al.,}{Cyburt
  et~al.}{2010}]{cyburt_jina_2010}
Cyburt R.~H.,  et~al., 2010, \mn@doi [The Astrophysical Journal Supplement
  Series] {10.1088/0067-0049/189/1/240}, 189, 240

\bibitem[\protect\citeauthoryear{Cyburt, Amthor, Heger, Johnson, Keek, Meisel,
  Schatz  \& Smith}{Cyburt et~al.}{2016}]{cyburt_dependence_2016}
Cyburt R.~H.,  Amthor A.~M.,  Heger A.,  Johnson E.,  Keek L.,  Meisel Z.,
  Schatz H.,   Smith K.,  2016, \mn@doi [The Astrophysical Journal]
  {10.3847/0004-637X/830/2/55}, 830, 55

\bibitem[\protect\citeauthoryear{Dorman \& Arnaud}{Dorman \&
  Arnaud}{2001}]{dorman_redesign_2001}
Dorman B.,  Arnaud K.~A.,  2001. p.~415, \url
  {http://adsabs.harvard.edu/abs/2001ASPC..238..415D}

\bibitem[\protect\citeauthoryear{Fujimoto}{Fujimoto}{1988}]{fujimoto_angular_1988}
Fujimoto M.~Y.,  1988, \mn@doi [The Astrophysical Journal] {10.1086/165955},
  324, 995

\bibitem[\protect\citeauthoryear{Fujimoto, Hanawa  \& Miyaji}{Fujimoto
  et~al.}{1981}]{fujimoto_shell_1981}
Fujimoto M.~Y.,  Hanawa T.,   Miyaji S.,  1981, \mn@doi [The Astrophysical
  Journal] {10.1086/159034}, 247, 267

\bibitem[\protect\citeauthoryear{Galloway}{Galloway}{2006}]{galloway_accretion-powered_2006}
Galloway D.~K.,  2006, \mn@doi [arXiv:astro-ph/0604345] {10.1063/1.2216602},
  840, 50

\bibitem[\protect\citeauthoryear{Galloway \& Cumming}{Galloway \&
  Cumming}{2006}]{galloway_helium-rich_2006}
Galloway D.~K.,  Cumming A.,  2006, \mn@doi [The Astrophysical Journal]
  {10.1086/507598}, 652, 559

\bibitem[\protect\citeauthoryear{Galloway, Muno, Hartman, Psaltis  \&
  Chakrabarty}{Galloway et~al.}{2008}]{galloway_thermonuclear_2008}
Galloway D.~K.,  Muno M.~P.,  Hartman J.~M.,  Psaltis D.,   Chakrabarty D.,
  2008, \mn@doi [The Astrophysical Journal Supplement Series] {10.1086/592044},
  179, 360

\bibitem[\protect\citeauthoryear{Galloway, Goodwin  \& Keek}{Galloway
  et~al.}{2017}]{galloway_thermonuclear_2017}
Galloway D.~K.,  Goodwin A.~J.,   Keek L.,  2017, \mn@doi [Publications of the
  Astronomical Society of Australia] {10.1017/pasa.2017.12}, 34, e019

\bibitem[\protect\citeauthoryear{Grindlay, Gursky, Schnopper, Parsignault,
  Heise, Brinkman  \& Schrijver}{Grindlay
  et~al.}{1976}]{grindlay_discovery_1976}
Grindlay J.,  Gursky H.,  Schnopper H.,  Parsignault D.~R.,  Heise J.,
  Brinkman A.~C.,   Schrijver J.,  1976, \mn@doi [The Astrophysical Journal
  Letters] {10.1086/182105}, 205, L127

\bibitem[\protect\citeauthoryear{Hartman et~al.,}{Hartman
  et~al.}{2008}]{hartman_long-term_2008}
Hartman J.~M.,  et~al., 2008, \mn@doi [The Astrophysical Journal]
  {10.1086/527461}, 675, 1468

\bibitem[\protect\citeauthoryear{Hartman, Watts  \& Chakrabarty}{Hartman
  et~al.}{2009}]{hartman_luminosity_2009}
Hartman J.~M.,  Watts A.~L.,   Chakrabarty D.,  2009, \mn@doi [The
  Astrophysical Journal] {10.1088/0004-637X/697/2/2102}, 697, 2102

\bibitem[\protect\citeauthoryear{He \& Keek}{He \&
  Keek}{2016}]{he_anisotropy_2016}
He C.-C.,  Keek L.,  2016, \mn@doi [The Astrophysical Journal]
  {10.3847/0004-637X/819/1/47}, 819, 47

\bibitem[\protect\citeauthoryear{Heger, Langer  \& Woosley}{Heger
  et~al.}{2000}]{heger_presupernova_2000}
Heger A.,  Langer N.,   Woosley S.~E.,  2000, \mn@doi [The Astrophysical
  Journal] {10.1086/308158}, 528, 368

\bibitem[\protect\citeauthoryear{Heger, Cumming, Galloway  \& Woosley}{Heger
  et~al.}{2007}]{heger_models_2007}
Heger A.,  Cumming A.,  Galloway D.~K.,   Woosley S.~E.,  2007, \mn@doi [The
  Astrophysical Journal Letters] {10.1086/525522}, 671, L141

\bibitem[\protect\citeauthoryear{Ibragimov \& Poutanen}{Ibragimov \&
  Poutanen}{2009}]{ibragimov_accreting_2009}
Ibragimov A.,  Poutanen J.,  2009, \mn@doi [Monthly Notices of the Royal
  Astronomical Society] {10.1111/j.1365-2966.2009.15477.x}, 400, 492

\bibitem[\protect\citeauthoryear{Jahoda, Swank, Giles, Stark, Strohmayer, Zhang
   \& Morgan}{Jahoda et~al.}{1996}]{jahoda_-orbit_1996}
Jahoda K.,  Swank J.~H.,  Giles A.~B.,  Stark M.~J.,  Strohmayer T.,  Zhang
  W.~W.,   Morgan E.~H.,  1996. International Society for Optics and Photonics,
  pp 59--71, \mn@doi{10.1117/12.256034}

\bibitem[\protect\citeauthoryear{Joss}{Joss}{1977}]{joss_x-ray_1977}
Joss P.~C.,  1977, \mn@doi [Nature] {10.1038/270310a0}, 270, 310

\bibitem[\protect\citeauthoryear{Keek \& Heger}{Keek \&
  Heger}{2011}]{keek_multi-zone_2011}
Keek L.,  Heger A.,  2011, \mn@doi [The Astrophysical Journal]
  {10.1088/0004-637X/743/2/189}, 743, 189

\bibitem[\protect\citeauthoryear{Kuulkers, den Hartog, in't Zand, Verbunt,
  Harris  \& Cocchi}{Kuulkers et~al.}{2003}]{kuulkers_photospheric_2003}
Kuulkers E.,  den Hartog P.~R.,  in't Zand J. J.~M.,  Verbunt F. W.~M.,  Harris
  W.~E.,   Cocchi M.,  2003, \mn@doi [Astronomy and Astrophysics]
  {10.1051/0004-6361:20021781}, 399, 663

\bibitem[\protect\citeauthoryear{Lampe, Heger  \& Galloway}{Lampe
  et~al.}{2016}]{lampe_influence_2016}
Lampe N.,  Heger A.,   Galloway D.~K.,  2016, \mn@doi [The Astrophysical
  Journal] {10.3847/0004-637X/819/1/46}, 819, 46

\bibitem[\protect\citeauthoryear{Levine, Bradt, Cui, Jernigan, Morgan,
  Remillard, Shirey  \& Smith}{Levine et~al.}{1996}]{levine_first_1996}
Levine A.~M.,  Bradt H.,  Cui W.,  Jernigan J.~G.,  Morgan E.~H.,  Remillard
  R.,  Shirey R.~E.,   Smith D.~A.,  1996, \mn@doi [The Astrophysical Journal
  Letters] {10.1086/310260}, 469, L33

\bibitem[\protect\citeauthoryear{Lewin, Vacca  \& Basinska}{Lewin
  et~al.}{1984}]{lewin_precursors_1984}
Lewin W. H.~G.,  Vacca W.~D.,   Basinska E.~M.,  1984, \mn@doi [The
  Astrophysical Journal] {10.1086/184202}, 277, L57

\bibitem[\protect\citeauthoryear{Lewin, Paradijs  \& Taam}{Lewin
  et~al.}{1993}]{lewin_x-ray_1993}
Lewin W. H.~G.,  Paradijs J.~V.,   Taam R.~E.,  1993, \mn@doi [Space Science
  Reviews] {10.1007/BF00196124}, 62, 223

\bibitem[\protect\citeauthoryear{Patruno \& Watts}{Patruno \&
  Watts}{2012}]{patruno_accreting_2012}
Patruno A.,  Watts A.~L.,  2012, preprint, 1206, arXiv:1206.2727

\bibitem[\protect\citeauthoryear{Patruno, Bult, Gopakumar, Hartman, Wijnands,
  van~der Klis  \& Chakrabarty}{Patruno
  et~al.}{2012}]{patruno_accelerated_2012}
Patruno A.,  Bult P.,  Gopakumar A.,  Hartman J.~M.,  Wijnands R.,  van~der
  Klis M.,   Chakrabarty D.,  2012, \mn@doi [The Astrophysical Journal Letters]
  {10.1088/2041-8205/746/2/L27}, 746, L27

\bibitem[\protect\citeauthoryear{Patruno et~al.,}{Patruno
  et~al.}{2017}]{patruno_radio_2017}
Patruno A.,  et~al., 2017, \mn@doi [The Astrophysical Journal]
  {10.3847/1538-4357/aa6f5b}, 841, 98

\bibitem[\protect\citeauthoryear{Rauscher, Heger, Hoffman  \& Woosley}{Rauscher
  et~al.}{2002}]{rauscher_nucleosynthesis_2002}
Rauscher T.,  Heger A.,  Hoffman R.~D.,   Woosley S.~E.,  2002, \mn@doi [The
  Astrophysical Journal] {10.1086/341728}, 576, 323

\bibitem[\protect\citeauthoryear{Strohmayer \& Bildsten}{Strohmayer \&
  Bildsten}{2003}]{strohmayer_new_2003}
Strohmayer T.,  Bildsten L.,  2003, arXiv:astro-ph/0301544

\bibitem[\protect\citeauthoryear{Tawara et~al.,}{Tawara
  et~al.}{1984}]{tawara_very_1984}
Tawara Y.,  et~al., 1984, \mn@doi [The Astrophysical Journal] {10.1086/184184},
  276, L41

\bibitem[\protect\citeauthoryear{Ubertini, Bazzano, Cocchi, Natalucci, Heise,
  Muller  \& in~'t Zand}{Ubertini et~al.}{1999}]{ubertini_bursts_1999}
Ubertini P.,  Bazzano A.,  Cocchi M.,  Natalucci L.,  Heise J.,  Muller J.~M.,
   in~'t Zand J. J.~M.,  1999, \mn@doi [The Astrophysical Journal Letters]
  {10.1086/311933}, 514, L27

\bibitem[\protect\citeauthoryear{Wang et~al.,}{Wang
  et~al.}{2001}]{wang_optical_2001}
Wang Z.,  et~al., 2001, \mn@doi [The Astrophysical Journal Letters]
  {10.1086/338357}, 563, L61

\bibitem[\protect\citeauthoryear{Weaver, Zimmerman  \& Woosley}{Weaver
  et~al.}{1978}]{weaver_presupernova_1978}
Weaver T.~A.,  Zimmerman G.~B.,   Woosley S.~E.,  1978, \mn@doi [The
  Astrophysical Journal] {10.1086/156569}, 225, 1021

\bibitem[\protect\citeauthoryear{Wijnands}{Wijnands}{2004}]{wijnands_observational_2004}
Wijnands R.,  2004, \mn@doi [Nuclear Physics B Proceedings Supplements]
  {10.1016/j.nuclphysbps.2004.04.084}, 132, 496

\bibitem[\protect\citeauthoryear{Wijnands \& van~der Klis}{Wijnands \& van~der
  Klis}{1998}]{wijnands_millisecond_1998}
Wijnands R.,  van~der Klis M.,  1998, \mn@doi [Nature] {10.1038/28557}, 394,
  344

\bibitem[\protect\citeauthoryear{Woosley \& Taam}{Woosley \&
  Taam}{1976}]{woosley_-ray_1976}
Woosley S.~E.,  Taam R.~E.,  1976, \mn@doi [Nature] {10.1038/263101a0}, 263,
  101

\bibitem[\protect\citeauthoryear{Woosley et~al.,}{Woosley
  et~al.}{2004}]{woosley_models_2004}
Woosley S.~E.,  et~al., 2004, \mn@doi [The Astrophysical Journal Supplement
  Series] {10.1086/381533}, 151, 75

\bibitem[\protect\citeauthoryear{in~'t Zand, Heise, Muller, Bazzano, Cocchi,
  Natalucci  \& Ubertini}{in~'t Zand et~al.}{1998}]{in_t_zand_discovery_1998}
in~'t Zand J. J.~M.,  Heise J.,  Muller J.~M.,  Bazzano A.,  Cocchi M.,
  Natalucci L.,   Ubertini P.,  1998, Astronomy and Astrophysics, 331, L25

\makeatother
\end{thebibliography}






\bsp	
\label{lastpage}
\end{document}